\documentclass[12pt]{iopart}

\usepackage{amssymb}
\usepackage{latexsym}
\usepackage{url}
\usepackage{hyperref}
\usepackage{graphicx}
\usepackage{booktabs}
\usepackage[mathscr]{euscript}
\usepackage[dvipsnames,table,xcdraw]{xcolor}
\usepackage{colortbl}
\expandafter\let\csname equation*\endcsname\relax
\expandafter\let\csname endequation*\endcsname\relax
\usepackage{amsmath}
\usepackage{iopams}
\usepackage[justification=raggedright,singlelinecheck=false]{caption}
\usepackage{rotating}

\begin{document}

\title[Preprint]
{Unsupervised deep learning model for fast energy layer pre-selection of delivery-efficient proton arc therapy plan optimization of nasopharyngeal carcinoma}

\author{Bohan Yang$^{1,5\dag}$, Gang Liu$^{2\dag}$, Yang Zhong$^{6\dag}$, Rirao Dao$^{4}$, Yujia Qian$^{4}$, Ke Shi$^{3}$, Anke Tang$^{3}$, Yong Luo$^{3\ast}$,  Qi Kong$^{6\ast}$, Jingnan Liu$^{1\ast}$}

\address{%
$^{1}$ Research Center of GNSS, Wuhan University, Wuhan, China. \\
$^{2}$ Cancer Center, Union Hospital, Tongji Medical College, Huazhong University of Science and Technology, Wuhan, China.\\
$^{3}$ School of Computer Science, National Engineering Research Center for Multimedia Software and Hubei Key Laboratory of Multimedia and Network Communication Engineering, Wuhan University, Wuhan, China.\\
$^{4}$ School of physics and technology, Wuhan University, Wuhan, China.\\
$^{5}$ Electronic Information School, Wuhan University, Wuhan, China. \\
$^{6}$ Department of Radiation Oncology, Anhui No. 2 Provincial People's Hospital, Hefei, China. \\
$^{\dag}$ Contribute equally
$^{\ast}$ Corresponding authors
}
\ead{yluo180@gmail.com, kq911203@163.com, jnliu@whu.edu.cn}

\vspace{10pt}
\begin{indented}
\item[] \textbf{Keywords:} Proton arc therapy; spot scanning; beam delivery time; energy layer selection; deep learning
\end{indented}

\begin{abstract}
{\it Objective.} Proton arc therapy (PAT) is an emerging and promising modality in radiotherapy, offering improved dose distribution and treatment robustness over conventional intensity-modulated proton therapy. Yet, identifying the optimal energy layer (EL) sequence remains challenging due to the intensive computational demand and prolonged in treatment delivery time. This study proposes an unsupervised deep learning framework for fast EL pre-selection that minimizes EL switch (ELS) time while maintaining high plan quality. 
{\it Approach.} We introduce a novel data representation method—spot-count representation—which encodes the number of proton spots intersecting the target and organs at risk (OAR) in a matrix structured by sorted gantry angles and energy layers. This representation serves as the input of an U-Net style architecture, SPArc\textsubscript{deep learning} (SPArc\textsubscript{dl}), which is trained using a tri-objective function: maximizing spot-counts on target, minimizing spot-counts on OAR, and reducing ELS time. The model is evaluated on 35 nasopharyngeal cancer cases, and its performance is benchmarked against plans generated by SPArc\textsubscript{particle swarm}. 
{\it Main results.} SPArc\textsubscript{dl} produces EL pre-selection that significantly improves both plan quality and delivery efficiency. Compared to SPArc\textsubscript{particle swarm}, it enhances the conformity index by 0.1 ($p < 0.01$), reduces the homogeneity index by 0.71 ($p < 0.01$), lowers the brainstem mean dose by 0.25 ($p < 0.01$), and shortens the ELS time by 37.2\% ($p < 0.01$). The results unintentionally reveal employing unchanged ELS is more time-wise efficient than descended ELS. SPArc\textsubscript{dl}'s inference time is within 1 second. However, SPArc\textsubscript{dl} plan demonstrates limitation in robustness. 
{\it Significance.} The proposed spot-count representation lays a foundation for incorporating unsupervised deep learning approaches into EL pre-selection task. SPArc\textsubscript{dl} is a fast tool for generating high-quality PAT plans by strategically pre-selecting EL to reduce delivery time while maintaining excellent dosimetric performance.
\end{abstract}

\section{Introduction}
Proton arc therapy (PAT) is an advanced and promising radiation therapy modality that combines the physical advantages of proton therapy with continuous rotational beam delivery. 
Since its initial introduction by Sandison et al.~\cite{sandison1997phantom} in 1997, PAT has seen renewed interest and technological advancement with the emergence of pencil beam scanning (PBS) and the development of Spot-Scanning Proton Arc Therapy (SPArc) by Ding et al.~\cite{ding2016spot} in 2016. Several studies have suggested that, in comparison to traditional methods like intensity-modulated proton therapy (IMPT), PAT may provide improved dose distribution~\cite{liu2021lung}, greater treatment robustness~\cite{liu2020improve, chang2020feasibility}, enhanced radio-biological effectiveness~\cite{bertolet2020proton, carabe2020radiobiological}, and a more streamlined treatment process~\cite{liu2020novel}. Clinical and simulation studies have since demonstrated PAT’s potential benefits across various tumor sites, including prostate, lung, breast, brain, and head and neck cancers~\cite{ding2018have, ding2019improving, liu2020improve, chang2020feasibility, li2018improve, seco2013proton, sanchez2016range}. 

A major approach to achieve the PAT planing optimization is to break it down into two steps: pre-selecting EL in advance, then using the selected EL to complete regular spot weight optimization~\cite{wuyckens2025proton}. In this approach, a significant challenge to the clinical adoption of PAT lies in the optimization of pre-selected energy layer (EL) sequences. In SPArc, the system is required to deliver protons at selected energy layers on each gantry angle. With hundreds of control points involved, frequent switching between energy layers can lead to substantial delays in treatment delivery. This is especially true for upward energy switches, which are time-intensive due to the magnetic hysteresis in the proton therapy system~\cite{gillin2010commissioning, zhu2010intensity, cao2014proton, van2015shortening, smith2009md}. Consequently, minimizing energy layer switch (ELS) time without compromising dosimetric quality has become a focus in PAT plan optimization.

Several EL sequence optimization approaches have been developed to address this challenge. Sanchez-Parcerisa et al.~\cite{sanchez2016range} proposes and compares four range optimization algorithms to select optimal mono- and bi-energetic beam energies for proton modulated arc therapy, aiming to minimize energy-switching time while maintaining dosimetric quality. However, this method shows constraints in performance when applied to intricate or irregular target geometries. The SPArc algorithm~\cite{ding2016spot} improves dose delivery by iteratively refining gantry angles and redistributing EL. As a subsequent method, SPArc\textsubscript{seq}~\cite{liu2020novel} reduces switching time by aligning energy sequences with the gantry’s rotational direction, and SPArc\textsubscript{particle swarm} (SPArc\textsubscript{ps}) \cite{qian2023novel} builds upon it by segmenting the arc into sectors based on water equivalent thickness (WET) and selecting EL that minimize switching time across control points. However, SPArc\textsubscript{ps} may need more objective functions on organs at risk (OARs) to optimize Head and Neck cancer, including nasopharyngeal carcinoma (NPC). Additionally, the ELO-SPAT framework~\cite{wuyckens2022treatment} integrates multiple objectives, including dose fidelity, energy sequencing constraints, and regularization terms, into a unified mathematical formulation that penalizes upward energy switches while exploiting downward transitions to enhance delivery efficiency. However, the optimization time reported in these studies~\cite{zhang2022energy, wuyckens2022treatment, ma2024machine}—ranging from several minutes to multiple hours—could pose a barrier to the clinical adoption of the framework. Based on the ground truth provided by modified ELO-SPAT, Ma et al.~\cite{ma2024machine} demonstrates the feasibility of using supervised machine learning method to integrate features in geometric, WET and beamlet and generate EL for each field. However, a supervised approach not only adds to the development workload due to ground truth generation but also constrains model performance to the quality of the pre-existing algorithms.


A gap in the literature is the lack of deep learning-based approaches, particularly using unsupervised learning, that can swiftly and directly predict optimal EL sequences. Deep learning has demonstrated strong potential in radiation oncology, offering high efficiency, automation, and generalization capacity~\cite{huynh2020artificial}. Yet, its application to EL pre-selection for PAT planning optimization remains largely unexplored. Unsupervised learning offers key advantages in this context: firstly it eliminates the workload of manually labeled data or algorithm-generated ground truths; secondly it removes the performance constrain brought by the quality of the pre-existing algorithm that generates the ground truth; thirdly, it enables researchers to directly embed domain knowledge into customized loss functions that reflect clinical priorities such as dose coverage, OAR sparing, and delivery efficiency.

In this study, we propose SPArc\textsubscript{deep learning} (SPArc\textsubscript{dl}), an unsupervised deep learning model for fast and effective energy layer pre-selection in PAT plan optimization. SPArc\textsubscript{dl} takes in data generated by spot-count representation, a novel data representation method specifically designed to support the application of deep learning models in EL pre-selection. This input data serves as a proxy of dose coverage and are analyzed by a U-Net style neural network. The model produces an EL probability distribution at all gantry angles and it is optimized using a tri-objective loss function that (1) maximizes target coverage, (2) minimizes OAR dose exposure, and (3) penalizes frequent and time-consuming EL switch (ELS).

We evaluate SPArc\textsubscript{dl} on 35 NPC cases and compare its performance with SPArc\textsubscript{ps}. Results show that SPArc\textsubscript{dl} significantly reduces ELS time—by 37.2\% ($p < 0.01$)—while enhancing conformity index (CI) by $0.1$ ($p < 0.01$), lowering homogeneity index (HI) by $0.71$ ($p < 0.01$), and reducing the mean dose of brainstem by $0.25$ ($p < 0.01$). By bypassing the need for supervised labels and offering rapid inference within $1$ second, SPArc\textsubscript{dl} demonstrates a viable path toward integrating deep learning into PAT planning workflows, supporting broad clinical adoption through enhanced efficiency and plan quality.

\section{Materials and methods}

\subsection{Dataset and spot-count data representation}
The treatment planning CT scans, acquired with a Philips Big Bore CT, of $35$ patients who has undergone NPC radiotherapy were collected from our collaborative hospital\footnote{Union Hospital, Tongji Medical College, Huazhong University of Science and Technology}. Our method for pre-selecting EL depends on the open-source software matRad \footnote{developed by Department of Medical Physics in Radiation Oncology, German Cancer Research Center-DKFZ} \cite{wieser2017development} which uses the CTs to generate the initial complete range of EL that longitudinally encompassed the entire treatment target at all gantry angles. 

The inclusion of diverse information—such as beamlet, WET and geometric-related features~\cite{ma2024machine}—may complicate the input structure of deep learning model. This study presents a novel data restructuring approach, termed spot-count representation, which constructs a data format specifically tailored to facilitate the application of deep learning models in EL pre-selection. The spot-count representation is founded on the assumption that the number of EL, denoted by $E$, and the number of gantry angles, denoted by $G$, are finite. Accordingly, we construct a matrix $M \in \mathbb{R}^{G \times E}$, where rows correspond to sorted gantry angles and columns to sorted EL, to record the number of proton spots intersecting a given anatomical structure. In this context, each element of $M$ represents spots count that intersect a specific structure $S$, organized by gantry angle and EL. For example, the number of spots intersecting CTV can be integrated into $M_{CTV}$. 

Building on the concept of matrix $M$, the number of spots intersecting one anatomical structure ($S_1$) but not another ($S_2$) can be represented by the element-wise subtraction $M_{S_1} - M_{S_2}$, resulting in a new matrix denoted as $M_{S_1-S_2}$. For instance, the number of spots that pass through the CTV while avoiding the brainstem is captured by the matrix $M_{CTV-Brainstem}$ ($M_{CTV-BRS}$). Since $M$ systematically records the number of spot intersections with anatomical structures, this data restructuring method is referred to as the spot-count representation.

\begin{figure}[ht]
    \centering
    \includegraphics[width=\textwidth]{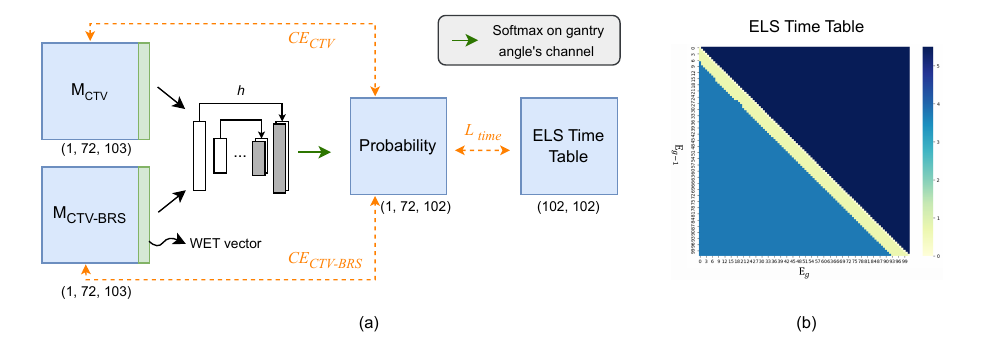}
    \caption{(a) SPArc\textsubscript{dl} model architecture. (b) ELS time table displays the time cost when the gantry angle transitions from $g-1$ to $g$.}
    \label{fig:arch}
\end{figure}

\subsection{Method}
We approximate the dose coverage of a volume as the number of spots intersection with the volume. Based on this approximation, matrix $M_{CTV}$ presents the dose coverage of CTV and $M_{CTV-BRS}$ presents dose covering CTV while dodging brainstem. Both matrices are formed and structured by gantry angles and EL. Previous method has found the WET value is key to mitigate the need for EL ascending and therefore improve delivery efficiency~\cite{qian2023novel}. As shown in \autoref{fig:arch}, the max value of WET of each gantry angle is calculated and fused with $M_{CTV}$ and $M_{CTV-BRS}$ by the gantry angle's channel. These two matrices are then concatenated into a 2-channels tensor and fed to the network, $h$. In our experiments, the input is of size $2 \times 72 \times 103$, but the network is not limited to this specific size. This input dimension happens to be similar to the dimensionality of an image, which is channel $\times$ height $\times$ width. In the past decade, many deep learning architecture have been developed to solve computer vision tasks. We use $h$, a network similar to the classic image segmentation method U-Net ~\cite{ronneberger2015u}, to analyze the spot counts of CTV and brainstem in the context of EL and gantry angles. 
The model consists of encoder and decoder sections with skip connection. We use convolution layers with a kernel size of $3$ and a stride of $1$ in both the encoder and decoder stages. Each convolutional layer is followed by a LeakyReLU activation function with a negative slope of $0.2$. The convolutional layers capture hierarchical spot-count distribution features of CTV and brainstem to estimate the optimal EL. The encoder employs average pooling to reduce the spatial dimensions in half at each level. During the decoding stage, we alternate between upsampling, applying convolutions, and concatenating skip connections that transfer features learned in the encoding stages directly to the layers responsible for generating the EL selection. The spatial resolution of the encoder levels is one half of the previous level, and that of the decoder levels is twice of the previous level. Four levels in total are included in the network with depth of $16$, $32$, $64$, and $128$. We set $1$ as the channel number of network's output which is of size $2 \times 72 \times 102$, and perform Softmax along the gantry angle's direction to obtain a probability tensor $P \in \mathbb{R}^{G \times EL}$. At inference stage, the EL corresponding to the highest probability of each gantry angle is selected as the predicted EL for that angle. We name this model, including the loss function, SPArc\textsubscript{dl}.

The dataset is divided randomly into $5$ folds for cross-validation, and the reported results represent the mean validation performance across all folds. The proposed model is trained end-to-end, optimized with the AdamW optimizer\cite{loshchilov2017decoupled} using a learning rate of $10^{-5}$ and a batch size of $16$. Training is conducted for $1000$ epochs. The model is implemented in PyTorch\cite{paszke2019pytorch} and trained on a single NVIDIA GeForce 3090 GPU, and inference is run on 11th Gen Intel(R) i7 CPU. 

\subsection{Unsupervised loss function}
Spot count is used as a proxy for dose coverage in SPArc\textsubscript{dl}. The three objectives of SPArc\textsubscript{dl} are to (1) maximize the dose coverage, or spot count, on CTV; (2) minimize the dose coverage, or spot count, on brainstem; and (3) reduce the ELS time across all gantry angles. The EL pre-selection task can be formulated as an EL classification problem at all angles. As a popular loss function of classification problems, cross entropy provides a well-founded, interpretable, and computationally efficient way to measures the difference between two probability distributions. In this paper, the in-used cross entropy loss is formulated in Eq.\eqref{eq:ce}.

\begin{equation}
  CE(M, P) = -\frac{1}{NG}\Sigma_{n=1}^N\Sigma_{g=1}^G\Sigma_{el=1}^{EL}(M_{g,el}\cdot log(P_{g,el})),
  \label{eq:ce}
\end{equation}
where $CE$ denotes cross entropy, $N$ denotes the case number of a batch, $EL$ denotes the number of energy layers, $G$ denotes the number of gantry angles, $M_{g,el}$ denotes the true labels for energy $el$ for angle $g$, and $P_{g,el}$ is the predicted probability for energy $el$ for angle $g$. 

Specifically, to achieve objective (1), the first part of the loss function of Eq.\eqref{eq:loss_func}, i.e. $CE(b(M_{CTV}), P)$, aims to shorten the distance between probability distribution of $P_g$ and $b(M_{CTV_g})$, so as to select the EL at angle $g$ that maximize the number of spots, or dose coverage, $g \in G$. $b$ is binarized-max function, which set the maximum value, or the maximum spot count, of each gantry angle as $1$ and other values as $0$. Like-wise, to achieve objective (2), the second part of Eq.\eqref{eq:loss_func}, i.e. $CE(b(M_{CTV-BRS}), P)$, is designed to maintain the coverage of CTV while reducing the brainstem dose exposure. The third part of Eq.\eqref{eq:loss_func} is introduced to penalize the ELS time, which can be generated using the ELS time table ($TB_{els}$) and formularized as Eq.\eqref{eq:time_loss}. 

\begin{equation}
  L_{time} = \frac{1}{NG} \Sigma_{n=1}^N\Sigma_{g=1}^G T(P_{g-1}, P_{g}, TB_{els}),
  \label{eq:time_loss}
\end{equation}
where $T(P_{g-1}, P_{g}, TB_{els})$ is the extraction of ELS time when the gantry angle switches from $g-1$ to $g$ given the time table $TB_{els}$. \autoref{fig:arch} (b) illustrates $TB_{els}$ where both the vertical (top to bottom) and horizontal (left to right) axes represent EL indices sorted in ascending order based on EL values. Both dimensions have a length of 102, reflecting the total of $102$ distinct EL presented in the dataset used. The cell value of $TB_{els}$ represents the ELS time between two adjacent gantry angles, $g_{-1}$ and $g$. It can be observed from \autoref{fig:arch} (b) the time cost of ascended ELS is higher than that of descended ELS. Note the ELS time is zero when EL is unchanged. The total loss function is

\begin{align}
  L_{total} &= \gamma \cdot CE_{CTV} + \zeta \cdot CE_{CTV-BRS} + \upsilon \cdot L_{time}    \\
  &= \gamma \cdot CE(b(M_{CTV}), P) + \zeta \cdot CE(b(M_{CTV-BRS}), P) + \upsilon \cdot L_{time}, 
  \label{eq:loss_func}
\end{align}
where $BRS$ denotes brainstem, $\gamma, \zeta$ and $\upsilon$ are hyper-parameters that can be tuned to ensure all components of the loss function are balanced. Through extensive experiments, we tune  $\gamma$, $\zeta$ and $\upsilon$ to be 0.5, 0.4 and 0.1, respectively.

\subsection{Evaluation}
A total of $35$ patients who underwent radiotherapy for NPC were retrospectively selected for evaluation by SPArc\textsubscript{ps} and SPArc\textsubscript{dl}. The performance of SPArc\textsubscript{dl} across all five validation datasets ($35$ cases) were aggregated for assessment. We calculate metrics mean for quantitative evaluation and select four representative cases for qualitative evaluation. Two structures, CTV and brainstem, are selected for evaluation. 
Full arc gantry angle selection was employed, with an angular sampling interval of $5$ degrees due to the constraints of computing power. Our energy selection method depended on matRad supplying the initial complete set of EL that span the entire treatment target longitudinally for every gantry angle. A spot spacing of $5$ mm was used in both the beam and lateral directions. Dose calculations were performed with a resolution of $3\times3\times3$ $mm^3$, and the voxel size of the CT images was $1.36\times1.36\times3$ $cm^3$. The prescription dose is $54$ Gy/CTV delivered in $30$ fractions, making $1.8$ Gy per fraction. 

\subsubsection{Plan quality evaluation}
The performance of SPArc\textsubscript{dl} in respect to plan quality and robustness of the PAT plan generated by the predicted EL is evaluated. 
The mean value of D\textsubscript{98}, D\textsubscript{2}, homogeneity index (HI), conformity index (CI), mean and max dose of brainstem are calculated for comparison between SPArc\textsubscript{dl} and SPArc\textsubscript{ps}. 

HI quantifies how uniformly the prescribed radiation dose is distributed within the target volume. A low HI value, ideally close to $0$, indicating a more homogeneous dose distribution, suggesting less dose variation within the target. Eq.\eqref{eq: HI} presents the formula of HI described in \cite{wu2003simultaneous}. 

\begin{equation}
  HI = \frac{D_2 - D_{98}}{D_{prescription}} * 100,
  \label{eq: HI}
\end{equation}
where D\textsubscript{2} and D\textsubscript{98} denotes the minimum dose received by 2\% and 98\% of the CTV, and D\textsubscript{prescription} refers to the prescription dose. CI measures how well the high-dose region conforms to the shape and size of the target volume. A CI value close to $1$ indicates a well-conformed treatment plan. Eq.\eqref{eq: CI} presents the formula of CI described in \cite{van1997conformation}. 

\begin{equation}
  CI = \frac{TV_{95}}{TV} \times \frac{TV_{95}}{V_{95}},
  \label{eq: CI}
\end{equation}
where TV\textsubscript{95} denotes the portion of the target volume receiving $95\%$ of the prescription dose, V\textsubscript{95} represents the total volume receiving at least $95\%$ of the prescription dose, and TV refers to the entire target volume. The mean and max dose of brainstem are reported to evaluate the OAR sparing of the proposed method. A lower amount of dose to brainstem represents a better brainstem protection in PAT. The qualitative evaluation to visualize plan quality is presented by the CT scans with dose distribution. 

A plan of a previous machine learning approach for EL prediction \cite{ma2024machine} reported a limited dose conformity index with regard to relatively small tumor sizes. Therefore, the plan quality metrics generated by the proposed deep learning approach is also assessed in respect to gross tumor volume (GTV). 

\subsubsection{EL switch time and model efficiency analysis}
A commonly emphasized objective in existing studies is the reduction of treatment delivery time \cite{wuyckens2025proton, qian2023novel, liu2020novel, ma2024machine, zhang2022energy}. The duration of arc delivery is influenced by various machine-specific factors, such as the number and order of EL, spot positions, and gantry angles. Among these, ELS time often contributes the most to overall delivery time, especially when the number of energy layers and energy switch-ups is not minimized \cite{wuyckens2025proton, qian2023novel, liu2020novel}. The ELS time is evaluated and analyzed for both SPArc\textsubscript{ps} and SPArc\textsubscript{dl}. The efficiency to generate EL pre-selection, or the inference time of SPArc\textsubscript{dl}, is also reported. 

\subsubsection{Robustness evaluation}
The robustness of the generated plan is assessed by dose volume histograms (DVHs) with bands and Root-Mean-Square Dose-Volume Histograms (RVH). They effectively capture the sensitivity of dose distributions to uncertainties such as patient setup errors and proton range variations. DVH bands illustrate the variability in dose coverage across multiple uncertainty scenarios, where narrower bands indicate greater dose stability or a more robust plan. RVH complements this by quantifying dose variation using the root-mean-square deviation across scenarios. A smaller RVH AUC reflects a more consistent and resilient dose distribution under uncertainty.

\begin{table}[htbp]
    \caption{The mean dosimetric performance comparison of 35 cases of SPArc\textsubscript{ps} and SPArc\textsubscript{dl}; D\textsubscript{98}: the minimum doses received by $98\%$ of the CTV; D\textsubscript{2}: the minimum doses received by $2\%$ of the CTV; CI: CTV conformity; HI: CTV homogeneity; Mean and Max: the mean dose and max dose of brainstem. }
    
    \label{tb:dosimetric}
    \centering
    \begin{tabular}{cccc}
        \toprule
        Structure	& Type   & SPArc\textsubscript{ps} & SPArc\textsubscript{dl} (\textbf{Ours})	\\
        \midrule
        CTV           &HI                       &0.97$\pm$0.57   	&0.26$\pm$0.04   \\
                      &CI                       &0.60$\pm$0.10   	  &0.70$\pm$0.02   \\
                      &D\textsubscript{2}       &1.99$\pm$0.10      &1.86$\pm$0.01   \\
                      &D\textsubscript{98}      &1.46$\pm$0.22   	  &1.72$\pm$0.01   \\
        Brainstem     &Max                      &1.91$\pm$0.12      &1.83$\pm$0.10 	 \\
                      &Mean                     &0.79$\pm$0.21   	  &0.54$\pm$0.16   \\
        \bottomrule
    \end{tabular}
\end{table}

\begin{figure}[ht]
    \centering
    \includegraphics[width=\textwidth]{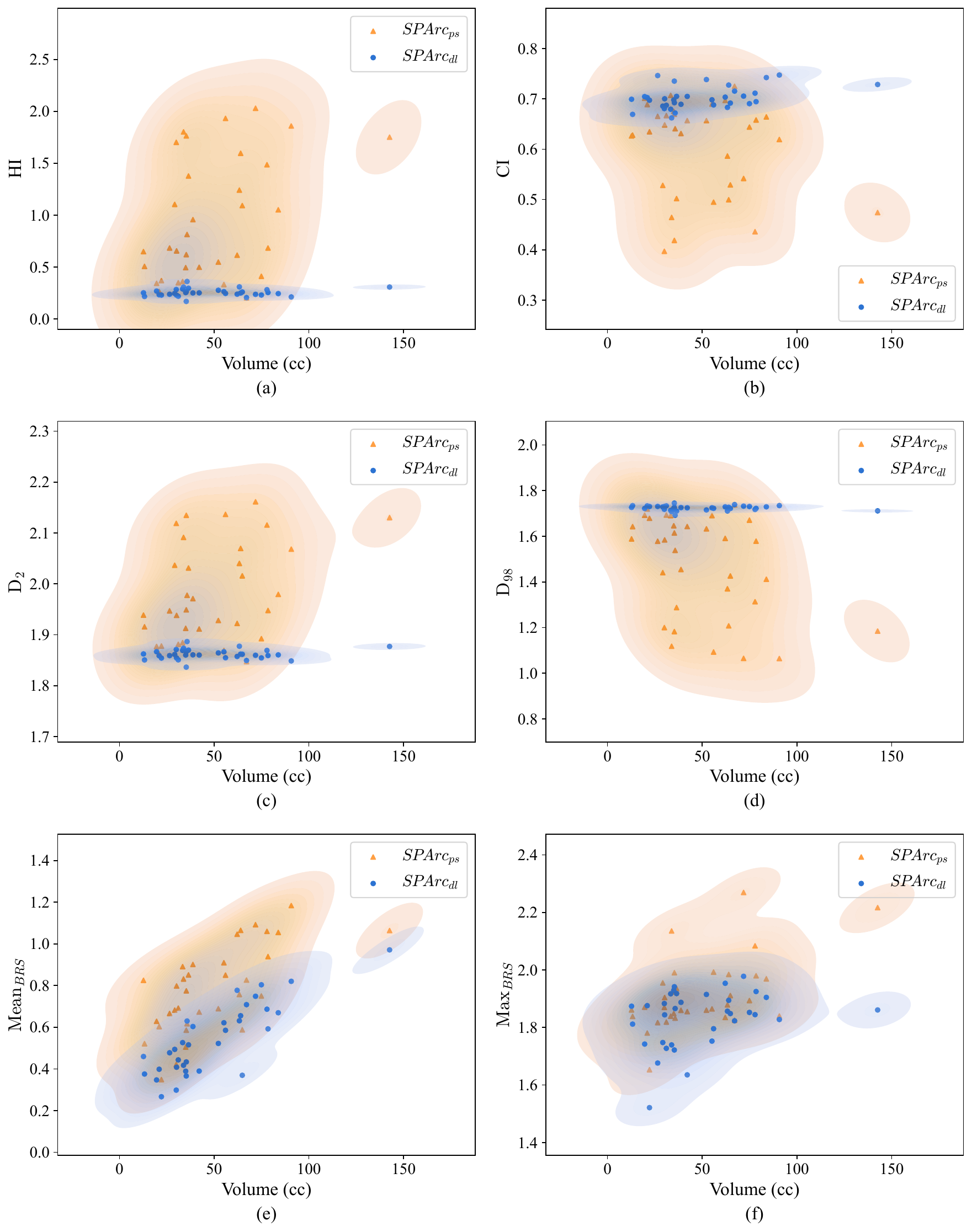}
    \caption{Comparison of various metrics performance of $35$ cases between SPArc\textsubscript{dl} and SPArc\textsubscript{ps} plans. Each diagram shows the relationship between GTV volume and the respective metric, displaying kernel density estimate plots and scatter points for both methods.}
    \label{fig:gtv_metrics_comparison}
\end{figure}

\section{Result}

\subsection{Plan quality}
\autoref{tb:dosimetric} demonstrates the quantitative dosimetric performance comparison of SPArc\textsubscript{dl} and SPArc\textsubscript{ps}, including D\textsubscript{98}, D\textsubscript{2}, CI, HI, mean and max dose of brainstem. With regard to target coverage, the  CI of SPArc\textsubscript{dl} is 0.70, higher than that of the SPArc\textsubscript{ps} ($p < 0.01$), suggesting SPArc\textsubscript{dl} achieves better dose conformity to the target. Compared to SPArc\textsubscript{ps}, the D\textsubscript{98} value of SPArc\textsubscript{dl} increased from $1.46$ to $1.72$ ($p < 0.01$) and D\textsubscript{2} decreases from $1.99$ to $1.86$ ($p < 0.01$). SPArc\textsubscript{dl} plan presents a significantly ($p < 0.01$) lower average HI compared to the SPArc\textsubscript{ps}, indicates more homogeneous dose distribution of SPArc\textsubscript{dl}. With regard to OAR sparing, the max and mean dose of brainstem (BRS) is reported in \autoref{tb:dosimetric}. Both metrics of BRS are lower in SPArc\textsubscript{dl} plan, with a difference of $0.08$  ($p < 0.01$) in max BRS dose and $0.25$ ($p < 0.01$) in mean BRS dose, suggeting improved sparing and better protection of BRS by SPArc\textsubscript{dl}. 

\begin{figure}[htbp]
    \centering
    \rotatebox[origin=c]{90}{\includegraphics[scale=0.75]{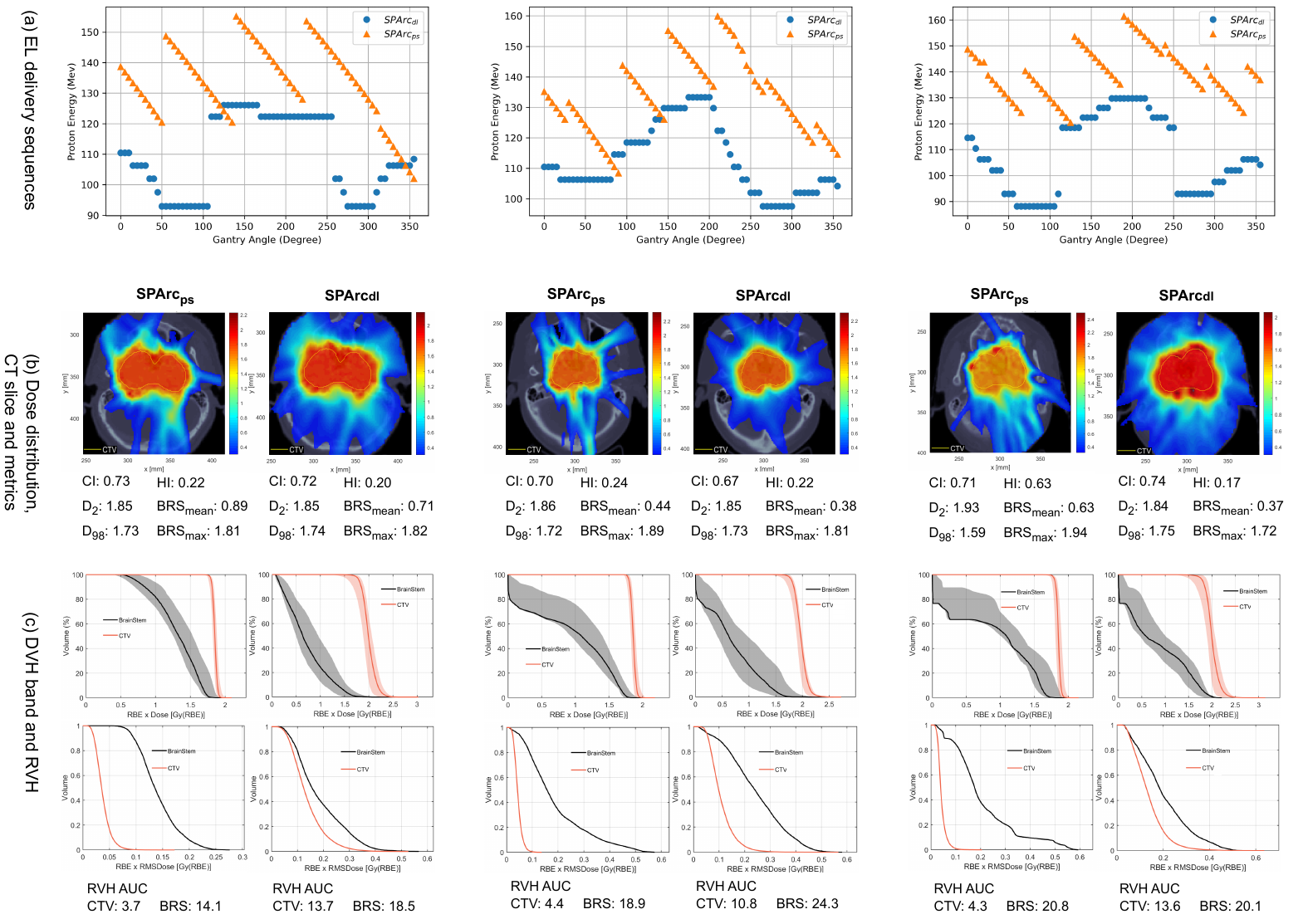}}
    \caption{Comparison of SPArc\textsubscript{dl} and SPArc\textsubscript{ps} plans. (a) The predicted EL delivery sequences of SPArc\textsubscript{ps} and SPArc\textsubscript{dl}. (b) The dose map overlaid on CT slice with six metrics for each case in nominal distribution. (c) The DVH band and RVH curve of the corresponding cases for robustness analysis. BRS stands for brainstem.}
    \label{fig:EL_seq_dose_metrics_robust}
\end{figure}

A more intuitive performance comparison of SPArc\textsubscript{dl} (blue dots) and SPArc\textsubscript{ps} (orange triangles) plans is demonstrated in \autoref{fig:gtv_metrics_comparison}, in which the distribution of CI, HI, D\textsubscript{98}, D\textsubscript{2}, mean and max dose of brainstem are presented against GTV using scatter plots and kernel density estimate plots. In plot (a), the HI values of SPArc\textsubscript{ps} plan range from about $0.4$ to $2$, with a dense cluster around from $0.4$ to $1$, while those of SPArc\textsubscript{dl} plan are lower and with a tighter cluster, mostly between $0.2$ and $0.4$, suggesting improved homogeneity of the proposed method. Also in plot (d), the D\textsubscript{98} values of SPArc\textsubscript{dl} plan are higher, mostly between $1.6$ and $1.8$, with a dense cluster around $1.7$ to $1.8$, indicating better minimum dose coverage. 
The CI values of SPArc\textsubscript{dl} are higher in plot (b), mostly between $0.6$ and $0.8$ with a dense cluster around $0.7$, while those of SPArc\textsubscript{ps} ranges from $0.5$ to $0.7$, showing better conformity of SPArc\textsubscript{dl} to the target volume. In plot (e) and (f), both Mean\textsubscript{BRS} and Max\textsubscript{BRS} of SPArc\textsubscript{dl} exhibits a lower range, suggesting better protection of brainstem. In particular, the Mean\textsubscript{BRS} values of both methods increase with the GTV, suggesting it is more difficult to spare OAR while GTV is large, which is consistent with expectations. The observations and findings of \autoref{fig:gtv_metrics_comparison} align with the insights of \autoref{tb:dosimetric}. Three cases are selected for qualitative evaluation in \autoref{fig:EL_seq_dose_metrics_robust}(b) which exhibit comparable CI and HI performance of SPArc\textsubscript{dl} and SPArc\textsubscript{ps}. Yet, the brainstem mean dose of SPArc\textsubscript{dl} is consistently lower than that of SPArc\textsubscript{ps}.


The relatively denser cluster generated by SPArc\textsubscript{dl} in the six plots of \autoref{fig:gtv_metrics_comparison} as well as the lower stardard deviation of SPArc\textsubscript{dl}'s performance arocss the six metrics in \autoref{tb:dosimetric} indicates greater consistency of SPArc\textsubscript{dl}'s performance. This suggests that SPArc\textsubscript{dl} not only improves the average dosimetric performance but also ensures more consistent and reproducible results in EL pre-selection.

\subsection{EL switch time and delivery sequence}

\autoref{fig:EL_seq_dose_metrics_robust}(a) demonstrates the prediction of EL delivery sequence of three cases by SPArc\textsubscript{dl} and SPArc\textsubscript{ps}. The predicted EL of SPArc\textsubscript{dl} are in the range between $90$ and $140$ Mev, while those of SPArc\textsubscript{ps} spans from approximately $110$ to $160$ Mev, suggesting a more conservative energy selection strategy by SPArc\textsubscript{dl}. The steeper diagonal trangle lines in the plots indicate SPArc\textsubscript{ps} utilizes a broader spectrum of energy layers, while many data points by SPArc\textsubscript{dl} cluster in horizontal segments, reflecting unchanged energy layers over multiple gantry angles. 


The performance of the model is also evaluated in terms of mean ELS time in \autoref{el_jump_switch_time}. While many previous works have found using descended ELS is more efficient than ascended ELS \cite{qian2023novel, liu2020novel, wang2025novel}, the results of SPArc\textsubscript{dl} unintentionally reveals using unchanged ELS may be more time-wise efficient than descended ELS. Unchanged ELS refers to the scenario in which the EL remains unchanged when the gantry angle transitions to the next position. In such cases, the ELS time is effectively zero, which is more efficient than that required by both ELS upward or downward transitions. \autoref{el_jump_switch_time} presents the mean ELS associated with ascended, descended and unchanged transition of SPArc\textsubscript{dl} and SPArc\textsubscript{ps}. While the mean ascended ELS of SPArc\textsubscript{dl} is $7.7$, slightly higher than that of SPArc\textsubscript{ps} ($6.6$), the mean unchanged ELS of SPArc\textsubscript{dl} is $53$, which is dramatically higher than that of SPArc\textsubscript{ps} or $0.3$. Compared to SPArc\textsubscript{ps}, the number of descended ELS of SPArc\textsubscript{dl} has vastly shifted to the unchanged ELS. The notable advantage of SPArc\textsubscript{dl} lies in its higher frequency of unchanged ELS. This advantage not only offsets the the cost of slightly higher EL in ascended transitions but also contributes to a gain of energy switch time of $30.1$ seconds, an $37.2\%$ reduction from that of SPArc\textsubscript{ps}. \autoref{fig:EL_seq_dose_metrics_robust}(a) demonstrates EL generated by SPArc\textsubscript{dl} of three cases, where many unchanged ELS can be observed.  

\begin{table}[htbp]
    \caption{The mean energy transitions and ELS time of 35 cases.}
    \label{el_jump_switch_time}
    \centering
    \begin{tabular}{cccc}
        \toprule
        & SPArc\textsubscript{ps}   & SPArc\textsubscript{dl}(Ours)	& Diff\\
        \midrule
        Number ascended ELS          &6.6     &7.7                  & +1.1 (16.7\%) \\
        Number descended ELS         &64.1   &10.3                  & -53.8 (-83.9\%) 	\\
        Number unchanged ELS         &0.3     &53.0                & +52.3 (176.7 times)	\\
        ELS time (sec) $\downarrow$      &81.1 $\pm$ 7.7   &51.0 $\pm$ 11.7   &\textbf{-30.1 (-37.2\%)}	\\
        \bottomrule
    \end{tabular}
\end{table}

\subsection{Robustness}
Comparative analysis of the DVH bands and RVH curves between SPArc\textsubscript{dl} and SPArc\textsubscript{ps} plans for two representative cases is presented in \autoref{fig:EL_seq_dose_metrics_robust}(c). While the DVH bandwidths for both CTV and brainstem demonstrate comparable characteristics between the two methods, notable differences emerge in their dose distribution patterns. SPArc\textsubscript{dl} exhibits higher-than-prescription doses in certain CTV regions, whereas SPArc\textsubscript{ps} maintains doses within prescribed limits. This disparity suggests increased dose coverage variability under uncertainties for SPArc\textsubscript{dl}, indicating its reduced plan robustness compared to SPArc\textsubscript{ps}. Analysis of RVH curves reveals that SPArc\textsubscript{dl} generates comparable area under curve (AUC) for brainstem, but larger area for CTV, further substantiating increased dose variation under uncertainty. The inferior robustness performance of SPArc\textsubscript{dl} can be attributed to the absence of robustness-specific objectives in its optimization framework, consequently limiting the network's capacity to improve robustness-related metrics such as DVH bands and RVH AUC. 



\subsection{Model Efficiency}
The mean inference time of SPArc\textsubscript{dl} for generating EL pre-selection for SPArc plan optimization is $0.05$ second on a Intel Core i7 CPU, representing a significant improvement in speed compared to the performance of SPArc\textsubscript{ps}, which is $13.5$ seconds ($p<0.01$) on the same CPU. This enhanced efficiency is primarily due to SPArc\textsubscript{dl}'s capability to simultaneously model the spot-organ and spot-target information across all gantry angles and produce EL pre-selection in a single forward pass.

\section{Discussion}
In this study, we proposed SPArc\textsubscript{dl}, a deep learning method to rapidly pre-select EL that maintain high treatment quality and significantly reduce ELS time in PAT, without relying on ground truth required by supervised approach. The proposed method is tested on $35$ NPC patient cases and evaluated against a published method SPArc\textsubscript{ps}. 

A novel data representation method, spot-count representation, involves the creation of matrices $M_{CTV}$ and $M_{CTV-BRS}$. Specifically, $M_{CTV}$ represents the number of proton spots intersecting the CTV, while $M_{CTV-BRS}$ indicates the spots intersecting CTV but avoiding the brainstem. Both matrices are structured by gantry angles and energy layers and serve as proxies for radiation dose coverage. SPArc\textsubscript{dl} processes these matrices along with WET vectors to determine the optimal energy layer for each gantry angle. The model employs a tri-objective loss function that maximizes target coverage by encouraging spot placement on CTV, minimizes organ-at-risk exposure by penalizing spots on the brainstem, and improves delivery efficiency by discouraging time-consuming energy layer switches. In terms of nominal dose distribution, SPArc\textsubscript{dl} achieves comparable or superior plan quality in CTV coverage and brainstem sparing compared to SPArc\textsubscript{ps}, with significantly reduced ELS time. However, further improvements in SPArc\textsubscript{dl}'s plan robustness are needed.

The contributions of this study is as follows. (1) To the best of our knowledge, the proposed SPArc\textsubscript{dl} is the first deep learning model solving EL pre-selection problem for PAT plan optimization. Compared to SPArc\textsubscript{ps}, SPArc\textsubscript{dl} demonstrates comparable or superior dosimetric performance, including dose conformity, homogeneity, and OAR sparing, while achieving greater consistency in results. (2) The experimental results of SPArc\textsubscript{dl} unintentionally revealed that deploying high frequency of unchanged ELS can be more time-wise efficient than high frequency of descended ELS. SPArc\textsubscript{dl} achieves improvements in delivery efficiency by leveraging unchanged ELS to significantly reduce ELS time. (3) This study introduces a novel data representation, termed spot-count representation, which is used as a proxy for dose coverage, enabling effective learning without direct dose calculations. Spot-count representation establishes an avenue to employ deep learning to the task of EL pre-selection. 
(4) The unsupervised approach of this study not only eliminates the complications associated with ground truth generation, but also leaves objective function open-ended where multiple objectives can be designed and simultaneously optimized, paving the way to deep learning-based multi-criteria optimization of SPArc plan optimization.
(5) SPArc\textsubscript{dl} enhances model inference speed by generating optimal EL sequences within 1 second, supporting fast clinical decision-making and planning workflows. 

While SPArc\textsubscript{dl} demonstrates promising results, its limitations merit careful consideration. First, the model currently lacks explicit mechanisms to address treatment uncertainties, particularly those arising from patient setup variations and intrafractional motion, which may impact the robustness of the selected EL in clinical settings. 
Second, computational resource limitations necessitate the sampling of full-arc gantry angles at 5-degree intervals instead of the more precise 2-degree intervals. Future research will focus on integrating robustness-aware strategies to improve both the accuracy and applicability of SPArc\textsubscript{dl} in real-world radiotherapy planning scenarios. 

\section{Conclusion}
In this study, we have demonstrated that SPArc\textsubscript{dl} achieves comparable or superior plan quality in nominal distribution while significantly improving delivery efficiency and accelerating inference speed in EL pre-selection task compared to the established SPArc\textsubscript{ps} method. The novel spot-count representation introduced here provides a strong foundation for implementing unsupervised deep learning methods in this task. An interesting finding is the experimental results suggest a high frequency of unchanged ELS deployments may be more efficient than a high frequency of descended ELS.
Even though SPArc\textsubscript{dl} has limitation regarding plan robustness, this research represents the first step towards integration of deep learning techniques in EL pre-selection for PAT plan optimization and may catalyze future developments in deep learning-based PAT plan optimization strategies.

\section*{Data availability statement}
The data cannot be made publicly available upon publication because no suitable repository exists for hosting data in this field of study. The data that support the findings of this study are available upon reasonable request from the authors.

\section*{Ethical statement}
This study was approved by the Institutional Review Board at the Tongji Medical College of Huazhong University of Science and Technology (ethics number: [2020] (0361)). All methods were performed in accordance with the relevant guidelines and regulations. Written informed consent was obtained from all authors and participants.

\section*{References}

\bibliographystyle{splncs04}
\bibliography{reference}

\end{document}